\documentclass[preprint,showpacs,preprintnumbers,amsmath,amssymb,nofootinbib]{revtex4}
\usepackage{bbm}
\usepackage{amsfonts}
\usepackage{booktabs}
\usepackage{mathrsfs}
\usepackage{epsfig}
\usepackage{graphicx}
\usepackage{dcolumn}
\usepackage{bm}
\usepackage{amsmath}
\usepackage{slashed}
\usepackage{subfigure}

\let\jnfont=\rm
\def\NPB#1,{{\jnfont Nucl.\ Phys.\ B }{\bf #1},}
\def\PLB#1,{{\jnfont Phys.\ Lett.\ B }{\bf #1},}
\def\EPJC#1,{{\jnfont Eur.\ Phys.\ Jour.\ C }{\bf #1},}
\def\PRD#1,{{\jnfont Phys.\ Rev.\ D }{\bf #1},}
\def\PRL#1,{{\jnfont Phys.\ Rev.\ Lett.\ }{\bf #1},}
\def\MPLA#1,{{\jnfont Mod.\ Phys.\ Lett.\ A }{\bf #1},}
\def\JPG#1,{{\jnfont J.\ Phys.\ G}{\bf #1},}
\def\CTP#1,{{\jnfont Commun.\ Theor.\ Phys.\ }{\bf #1},}
\def\ZPC#1,{{\jnfont Z.\ Phys.\ C }{\bf #1},}
\def\JHEP#1,{{\jnfont JHEP \ }{\bf #1},}
\def\Rv{\not{\hbox{\kern-1pt $R$}}}
\def\p{\not{\hbox{\kern-3pt $p$}}}

\begin{document}

\title{Complete one-loop effects of SUSY QCD in $b\bar{b}h$ production at the LHC
         under current experimental constraints}

\author{Ning Liu$^1$, Lei Wu$^2$, Pei Wen Wu$^2$, Jin Min Yang$^2$
        \\~ \vspace*{-0.3cm} }
\affiliation{ $^1$ Physics Department, Henan Normal University,
     Xinxiang 453007, China\\
$^2$ Institute of Theoretical Physics,
     Academia Sinica, Beijing 100190, China
     \vspace*{1.5cm}}

\begin{abstract}
Inspired by the recent LHC Higgs data and null search results of
supersymmetry (SUSY), we scan the parameter space of the Minimal
Supersymmetric Standard Model (MSSM) with relatively heavy
sparticles (1-3 TeV). Then in the parameter space allowed by current
collider experiments and dark matter detections, we calculate the
complete one-loop SUSY QCD corrections to $pp\to b\bar{b}h$ at the
LHC with $\sqrt{s}=14$ TeV and obtain the following observations:
(i) For the large values of $\tan\beta$ and low values of $m_{A}$,
the SUSY QCD effects can be quite large, which, however, have been
excluded by the latest results of LHC search for $H/A \to
\tau^{+}\tau^{-}$; (ii) For modest values of $\tan\beta$ and $m_{A}$
which so far survived all experimental constraints, the SUSY QCD
corrections can maximally reach about $-9\%$.

\end{abstract}
\pacs{}

\maketitle

\section{INTRODUCTION}

Very recently the ATLAS and CMS collaborations have independently
reported the observation of a Higgs-like resonance with a mass about
125 GeV \cite{recent-LHC}. At the same time, the CDF and D0 collaborations
have also updated their combined results for the Higgs searches in
$b\bar{b}$ channel, which support the LHC observation \cite{tevatron-higgs}.
Since in the Minimal Supersymmetric Standard Model (MSSM) a SM-like Higgs
boson is predicted with a mass below 130 GeV, the observation of
such a 125 GeV Higgs boson supports SUSY, albeit quite restrictive
on the parameter space of SUSY \cite{125-cmssm}.

Meanwhile, the direct searches for SUSY particles (sparticles)
have been performed at the LHC. Based on about 5 fb$^{-1}$ luminosity,
the ATLAS and CMS collaborations have reported null results and obtained
some bounds on the sparticle masses, which is about 1 TeV
for the gluino and first generation of squarks \cite{squark-gluino mass},
 330 GeV for the electroweak gauginos, 180 GeV for the sleptons
\cite{gaugino-slepton mass}, 465 GeV for the stops and 480 GeV for the
sbottoms \cite{stop-sbottom mass}.
These bounds indicate that SUSY may be heavier than expected
and the sparticles may be significantly heavier than the electroweak scale
\cite{effective-susy-1,effective-susy-2}.

In case that the sparticles are heavy and beyond the LHC scope of
direct production, search for the indirect SUSY effects via loop
corrections will be of great importance. Since the loop effects of
heavy sparticles are usually small, we should look for some
processes in which the heavy sparticles have residual loop effects.
One type of such processes are Higgs productions at the LHC, such as
the production of $tH^{-}$ and $h b \bar{b}$, in which the heavy
sparticles have sizable residual loop effects for a small value of
$m_{A}$ and a large value of $\tan\beta$ \cite{non-decoupling,bbh}
(when $m_{A}$ getting large, such effects will vanish). The reason
for these residual loop effects is that the couplings in the loops
are proportional to some SUSY mass parameters and can be enhanced by
the large values of $\tan\beta$.

In this note we focus on the production  of $h b\bar{b}$ at the LHC
and calculate the complete one-loop SUSY QCD corrections to this
process. As an important Higgs production channel for the MSSM, this
production has been studied in the literature \cite{bbh}, where the
residual SUSY QCD effects are found to be large (reach -40\% for
$\tan\beta=30$). We revisit this production for the following
reasons: Firstly, in the literature the SUSY QCD corrections to this
process are calculated only partially (only the corrections to the
$h b\bar{b}$ vertex have been considered). The complete one-loop
corrections involve pentagon Feynman diagrams, whose calculations
are rather complicated and have not been performed. Secondly, the
CMS collaboration has recently measured this channel and given
constraints on the plane of $\tan\beta$ versus $m_{A}$
\cite{hbb-exp}. Since the residual SUSY QCD effects in this
production is sensitive to the values of  $\tan\beta$ and $m_{A}$,
we should update the calculations by considering such new
constraints. Moreover, other experimental constraints, such as the
dark matter direct detection limits and the SM-like Higgs boson mass
around 125 GeV, are also rather restrictive and should be
considered. In this work, we consdier all current experimental
constraints to scan the MSSM parameter space and then in the allowed
parameter space we calculate the process $pp \to b\bar{b}h$ with the
complete one-loop SUSY QCD corrections.

 The paper is organized as follows. In Sec. II. we
 describe the calculations for the
process $pp \to b\bar{b}h$. In Sec.III we show numerical
results. Finally, we draw the conclusions in Sec. IV.

\section{The description of calculations}
In the MSSM the lighter CP-even Higgs mass ($m_{h}$) is smaller than
$M_{Z}$ at tree level but receives large corrections at the loop level.
The leading part of the corrections is from the stop sector and can be
expressed as \cite{mh-1loop}
\begin{eqnarray}
\Delta m^{2}_{h}(\tilde{t})\simeq\frac{3m^{4}_{t}}{2\pi^{2}v^{2}\sin^{2}\beta}[\log
\frac{m_{\tilde{t}_{1}}m_{\tilde{t}_{2}}}{m^{2}_{t}}
+\frac{X^{2}_{t}}{2m_{\tilde{t}_{1}}m_{\tilde{t}_{2}}}(1-\frac{X^{2}_{t}}
{6m_{\tilde{t}_{1}}m_{\tilde{t}_{2}}})]
\end{eqnarray}
where $X_{t}=A_{t}-\mu\cot\beta$ is the stop mixing parameter.
We see that a large stop mass or a large stop mixing parameter is needed to
increase $m_{h}$ to 125 GeV.
In our calculations we consider the collider constraints on the MSSM Higgs sector,
using the packages \textsf{FeynHiggs2.8.6} \cite{feynhiggs} and
\textsf{HiggsBounds-3.8.0}\cite{higgsbounds} to calculate the
observables in the Higgs sector and require them to satisfy the
constraints from  the LEP, Tevatron and LHC.

The SUSY QCD corrections to $hb\bar{b}$ production at the LHC
involve the sbottoms and gluino in the loops. The sbottom mass
matrix takes the form \cite{susyint}
\begin{equation}
M_{\tilde b}^2 =\left(\begin{array}{cc}
m_{{\tilde b}_L}^2& m_bX_b^\dag\\
 m_bX_b& m_{{\tilde b}_R}^2 \end{array} \right) \ ,
\end{equation}
where
\begin{eqnarray}
m_{{\tilde b}_L}^2 &=& m_{\tilde Q}^2+m_b^2-m_Z^2(\frac{1}{2}
-\frac{1}{3}\sin^2\theta_W)\cos(2\beta) \ , \\
m_{{\tilde b}_R}^2 &=& m_{\tilde D}^2+m_b^2 -\frac{1}{3}m_Z^2 \sin^2\theta_W\cos(2\beta) \ , \\
X_b&=& A_b-\mu\tan\beta \ ,
\end{eqnarray}
where $m_{\tilde Q}^2$ and $m_{\tilde{D}}^2$ are respectively the soft-breaking mass parameters
for the left-handed squark doublet $\tilde Q$ and the right-handed down squark
$\tilde D$, $A_b$ is the sbottom soft-breaking trilinear coupling and $\mu$ is
the SUSY-preserving bilinear coupling of the two Higgs doublets in the superpotential.
This mass matrix can be diagonalized by a unitary transformation
which rotates the weak eigenstates $\tilde b_{L,R}$ to the mass eigenstates
$\tilde b_{1,2}$,
\begin{eqnarray}
\left (\begin{array}{c} \tilde b_1 \\ \tilde b_2 \end{array} \right )
= \left ( \begin{array}{cc}\cos\theta_{\tilde{b}} &\sin\theta_{\tilde{b}} \\
-\sin\theta_{\tilde{b}} &\cos\theta_{\tilde{b}} \end{array} \right )
\left (\begin{array}{c} \tilde b_L \\ \tilde b_R \end{array} \right)
\end{eqnarray}
with the sbottom masses $m_{\tilde b_{1,2}}$ and the mixing angle $\theta_{\tilde{b}}$
determined by
\begin{eqnarray}
m_{\tilde b_{1,2}}&=&\frac{1}{2}\left[ m_{{\tilde b}_L}^2+m_{{\tilde b}_R}^2
\mp\sqrt{\left(m_{{\tilde b}_L}^2-m_{{\tilde b}_R}^2\right)^2 +4m_b^2X_b^2}\right] ,\\
\tan2\theta_{\tilde{b}} &=& \frac{2m_bX_b}{m_{{\tilde b}_L}^2-m_{{\tilde b}_R}^2} \ .
\end{eqnarray}

\begin{figure}[htbp]
\includegraphics[width=5in]{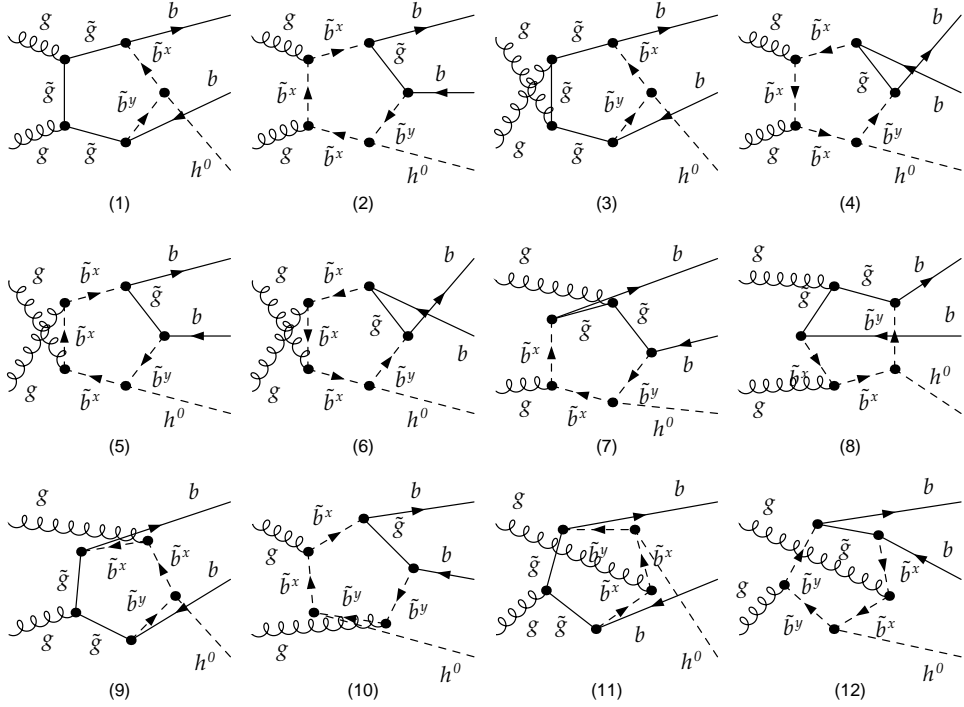}%
\vspace{-0.5cm} \caption{The pentagon diagrams for SUSY QCD
corrections to the subprocess of $gg \to b\bar{b}h$ at the LHC.}
\label{fig1}
\end{figure}
We produce the one-loop amplitudes with \textsf{FeynArts-3.5}
\cite{feynart} and use the \textsf{FormCalc-6.1} \cite{formcalc} to
simplify them and express the loop functions. The numerical
calculations are performed by using \textsf{LoopTools-2.2}
\cite{looptools}. In Fig.1 we display the representative pentagon
Feynman diagrams for the SUSY QCD corrections in the subprocesses
$gg \to b\bar{b}h$. Due to no massless particles in the loop, all
the Feynman diagrams with the gluino and sbottoms in the loops are
infrared (IR) finite.

We take the definitions of the scalar and tensor two-, three-, four-
and five-point integral functions presented in Ref.\cite{denner}.
For the calculation of the pentagon diagrams, we use
Passarino-Veltman method\cite{loop function} to reduce the N-point(N
$\leq$ 5) tensor functions to scalar integrals. Our programs have
been used to study the SUSY-QCD corrections to the process $pp \to
t\bar{t}Z^{0}$ at the LHC \cite{liu-ning} and have been checked with
Ref.\cite{ttz-sm} therein. In order to further validate the
calculation of the pentagon diagrams, we used our programs to
calculate the NLO QCD corrections to $pp \to t\bar{t}h$ in the SM at
the LHC and compared with the results in Ref.\cite{tth}. As shown in
Table I, our results argee with those in Ref.\cite{tth} very well.
In order to preserve supersymmetry, we adopt the constrained
differential renormalization (CDR) \cite{cdr1} to regulate the
ultraviolet divergence (UV) in the self-energy and vertex
corrections, which is equivalent to the dimensional reduction method
at one-loop level \cite{cdr2}.

\begin{table}[h]\caption{\label{tab1} The comparison of our numerical results for
the process $pp\to t\bar{t}h$ in the SM at the LHC with those in
Ref.\cite{tth}, where the LO and NLO QCD corrected cross sections
for different Higgs mass are listed with the relevant parameters and
the PDFs being the same as in Ref.\cite{tth}, i.e., $\mu_0 = (2 m_t
+ m_h)/2$, $m_t = 174$~GeV and the MRST PDFs.}
\begin{tabular}{|c|c|c|c|c|}
\hline $m_h$(GeV) & our $\sigma_{LO}$(fb) & our $\sigma_{NLO}$(fb) & $\sigma_{LO}$(fb) in \cite{tth} & $\sigma_{NLO}$(fb) in \cite{tth}   \\
\hline $120$      & 577.4(4)              & 701.3(13)              &  577.3(4)                 & 701.5(18)       \\
\hline $140$      & 373.6(2)              & 452.4(11)                &  373.4(3)               & 452.3(12)       \\
\hline $160$      & 251.3(4)              & 305.5(7)                &  251.6(2)                & 305.6(8)       \\
\hline
\end{tabular}
\end{table}

In our calculations, we assume a common SUSY mass $M_{SUSY}$ defined by
$M_{SUSY}=M_{\tilde{Q}}=M_{\tilde{U}}=M_{\tilde{D}}=M_{\tilde{g}}=A_t=A_b=\mu$.
We fix slepton mass parameters
$M_{\tilde{L}}=M_{\tilde{E}}=A_{\tau}=3$ TeV and scan the following
MSSM parameter regions:
\begin{eqnarray}
&&~~~~ 5 \le \tan\beta \le 60, \quad 90~\rm GeV \le M_A \le 350~\rm
GeV, \quad 1~\rm TeV \le M_{SUSY} \le 3~\rm TeV
\end{eqnarray}
In our scan we consider the following constraints on the parameter
space: (i) We require that the bounds for Higgs bosons from LEP,
Tevatron and LHC are satisfied and the mass of light CP-even Higgs
is in the region of 123 GeV$<m_{h}<$ 129 GeV; (ii) For the
constraints from flavor physics and electroweak precision tata, we
checked by using the package \textsf{susy$\_$flavor v2.0}
\cite{susy-flavor} that they are safely satisfied because we assume
relatively heavy sparticles. (iii) We consider the dark matter
constraints from the WMAP relic density and the direct detection
results by using the package \textsf{MicrOmega v2.4}
\cite{micromega}.

\section{Numerical Results}
Since the b-quark Yukawa coupling may receive large
radiative corrections in the MSSM, we use the running b-quark
mass ($m^{\overline{DR}}_{b}$) and use the method induced in
\cite{hollik} to absorb the MSSM corrections into the effective
 b-quark Yukawa couplings. But for the b-quark in the final state, we
take the pole mass to assure the correct on-shell behavior.

In our numerical calculations, we take the input parameters of the
SM as \cite{pdg}
\begin{eqnarray*}
m_t=172{\rm ~GeV},
~~m^{\overline{MS}}_{b}(m^{\overline{MS}}_{b})=4.19{\rm ~GeV},
~~m_{Z}=91.1876 {\rm ~GeV},~~\alpha(m_Z)=1/127.918
\end{eqnarray*}
Here $m^{\overline{MS}}_{b}(m^{\overline{MS}}_{b})$ is the
QCD-$\overline{MS}$ bottom-quark mass, which is related to
$m^{\overline{DR}}_{b}$ as
\begin{eqnarray}
m^{\overline{DR}}_{b}=m^{\overline{MS}}_{b}[1-\frac{\alpha_{s}}{3\pi}
-\frac{\alpha^{2}_{s}}{144\pi^{2}}(73-3n_f)]
\end{eqnarray}
where $n_f$ is the number of the active quark flavors. For the
strong coupling constant $\alpha_s(\mu)$, we take its 2-loop
evolution with QCD parameter $\Lambda^{n_{f}=5}=226{\rm ~MeV}$
and get $\alpha_s(m_Z)=0.118$. We use CTEQ6L1 and
CTEQ6M \cite{cteq} parton distribution functions (PDF) for the SM
tree level and SUSY QCD one-loop level computations, respectively.
The renormalization scale $\mu_R$ and factorization scale $\mu_F$ are
chosen to be $\mu_R=\mu_F=m_Z$. We numerically checked that all the
UV divergence in the loop corrections canceled.

\begin{figure}[htbp]
\includegraphics[width=4in,height=3.5in]{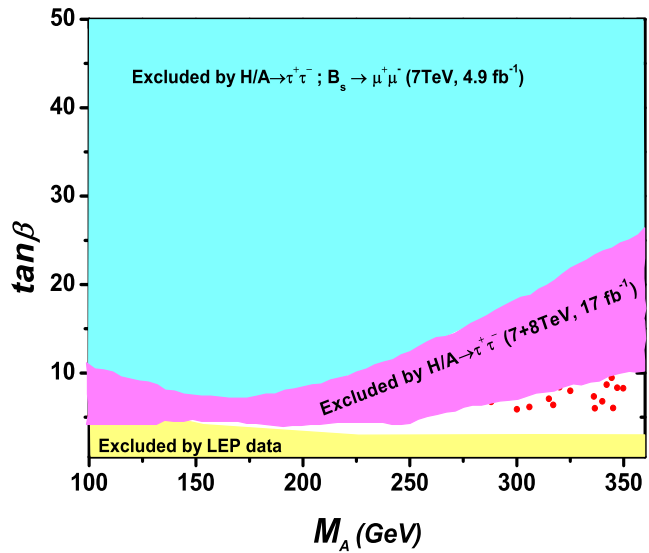}%
\vspace{-0.5cm} \caption{The parameters space satisfying constraints
(i)-(iii), projected on the planes of $\tan\beta$ versus $m_A$. The
blue region is excluded by the LHC data, in particular, by the
search of new particles decaying into $\tau^{+}\tau^{-}$ and the
measurement of $B_s \to \mu^{+}\mu^{-}$. The yellow region is
excluded by the non-observation of Higgs boson at LEP2. The red dots
represent the samples survied all the constraints.} \label{fig2}
\end{figure}

In Fig.2 we project the survived samples satisfying all the
experimental constraints on the planes of $\tan\beta$ and $M_h$
versus $m_A$ and also present the excluded regions. It can be seen
that a large part of the parameter space (the light blue region) has
been ruled out by the 7 TeV LHC data, in particular, by the search
of new particles decaying into $\tau^{+}\tau^{-}$ and the
measurement of $B_s \to \mu^{+}\mu^{-}$. For the small $\tan\beta$
and low and moderate $M_A$ region, it has been excluded by the
non-observation of Higgs boson at LEP2. We also note that, with the
very recently released 7+8 TeV LHC results of $H/A\to
\tau^{+}\tau^{-}$ based on $\mathcal{L}=17$
fb$^{-1}$\cite{cms-tautau}, the excluded lower limit on the plane of
$\tan\beta-M_A$ has been further pushed down and overlaps with the
one of LEP2 in the low $M_A$ case. Since we require the Higgs boson
mass to be in the range of 123-129 GeV indicated by the LHC
data(126$\pm$3 GeV), the parameter space that can correctly produce
the Higgs boson mass is highly constrained and situated in a region
with a modest $M_A$ ($M_A\gtrsim 200$ GeV) and a small $\tan\beta$
($6\lesssim\tan\beta\lesssim 12$). However, for other parts of the
parameters space, they produce a too heavy Higgs boson ($m_h>129$
GeV when $\tan\beta>12$) or a too light Higgs boson ($m_h<123$ GeV
when $\tan\beta<6$).

\begin{figure}[htbp]
\includegraphics[width=5in,height=4.5in]{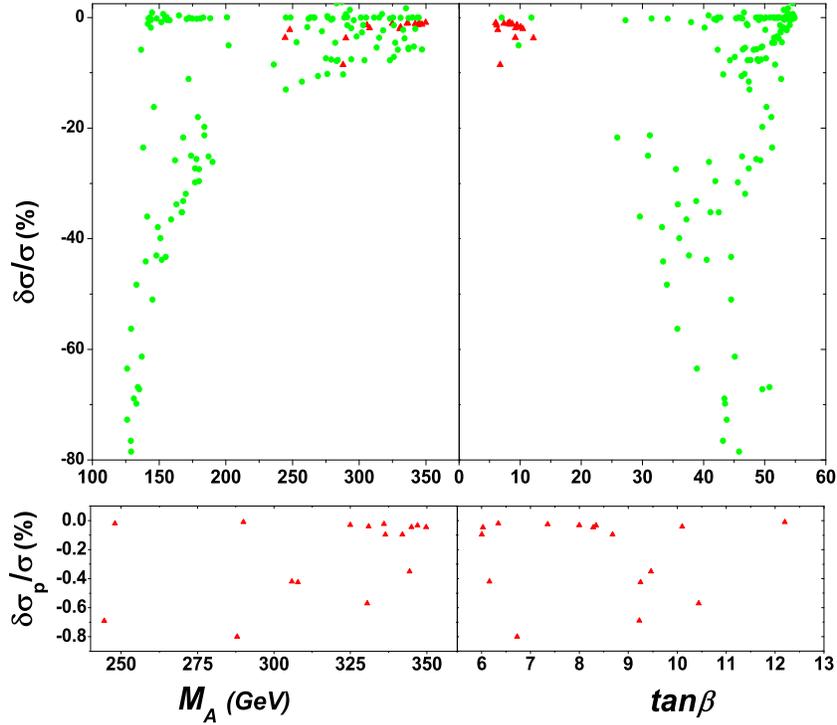}%
\vspace{-0.5cm} \caption{The upper panel shows the complete one-loop
SUSY QCD corrections($\delta\sigma/\sigma$) versus $M_A$ and
$\tan\beta$ for the samples satisfying(red triangles) or not
satisfying(green dots) the LHC constraints. The lower panel shows
the SUSY-QCD corrections from the contributions of pentagon
diagrams($\delta\sigma_{p}/\sigma$) for the red triangle samples.}
\label{fig2}
\end{figure}

In Fig.3 we present seperately the pentagon diagram contribution
(lower panel) and the total SUSY-QCD contribution (upper panel) for
the surviving samples. In order to show the influence of the recent LHC data,
we present the complete one-loop SUSY QCD
corrections for the samples satisfying or not satisfying the LHC constraints.
We can see that
the complete SUSY QCD corrections will be significant for the
samples which have a large $\tan\beta$ and a low value of $m_{A}$.
This can be understood by the contribution to the effective b-quark
Yukawa coupling after integration of the heavy sparticles, which is
$\delta \bar{y}_{hb\bar{b}}=-\frac{g \alpha_s
m^{\overline{MS}}_{b}\sin\alpha}{6\pi m_W
\cos\beta}(\frac{M_{\tilde{g}}\mu}{M^{2}_{SUSY}})[\tan\beta+\cot\alpha]$
\cite{effective-susy-2}. Since we assume
$M_{SUSY}=M_{\tilde{g}}=\mu$, the Yukawa coupling will be
independent of the sparticle masses and be greatly enhanced by a
large $\tan\beta$. However, it should be noted that these samples
will lead to the excess of the production rate of $pp \to H/A \to
\tau^{+}\tau^{-}$ and thus have been excluded by the current
measurements. With the increase of $m_{A}$ and the decrease of
$\tan\beta$, the corrections drop rapidly and approach zero in the
decoupling limit. The main reason is that $\delta
\bar{y}_{hb\bar{b}}$ can be heavily reduced by the cancellation
between $\tan\beta$ and $\cos\alpha$, which have a relation as
$\cot\alpha\simeq-\tan\beta-2m^{2}_{Z}\tan\beta\cos^{2}\beta/m^{2}_{A}$
for a large $m_A$. From the lower panel of Fig.3, we see that
contributions of those pentagon diagrams are small and maximally
reach about 0.8\%. This is because in those pentagon diagrams
the Higgs boson only couples to sbottoms while the $hb\bar{b}$ vertex
does not appear. So the large residual loop effects in the $hb\bar{b}$ vertex
are absent in the pentagon diagrams.
For the samples which survived all the constraints, the
complete SUSY QCD corrections can only reach about $-9\%$ at the LHC
with $\sqrt{s}=14$ TeV. Detecting such a size of SUSY QCD effects
may be challenging in the future measurement of the process $pp \to
b \bar{b} h$ \cite{gunion}.

\section{conclusion}
In this work, we calculated the complete one-loop SUSY QCD corrections
to the process $pp\to b\bar{b}h$ at the LHC with $\sqrt{s}=14$ TeV.
We found that the large SUSY QCD corrections in the non-decoupling
regime with a large $\tan\beta$ and a low $m_{A}$
has been excluded by the latest results of LHC non-standard Higgs
searches. For the survived decoupling regime which have
modest values of $\tan\beta$ and $m_{A}$, the
SUSY QCD corrections can maximally reach $-9\%$.

\section*{Acknowledgement}
This work was supported in part by the National Natural Science
Foundation of China (NNSFC) under grant Nos. 10821504 and 11135003,
by the Project of Knowledge Innovation Program (PKIP) of Chinese
Academy of Sciences under grant No. KJCX2.YW.W10. and by the Startup
Foundation for Doctors of Henan Normal University under contract
No.11112.

\end{document}